\documentclass[aps,prb,twocolumn,superscriptaddress,citeautoscript]{revtex4-1}
\usepackage{graphicx}
\usepackage{amsmath, amsfonts}
\usepackage{amssymb}
\usepackage[colorlinks=true,linkcolor=blue,citecolor=blue]{hyperref}
\usepackage{color}
\usepackage[utf8]{inputenc}

\begin{document}
	
	\author{T\'imea N\'ora T\"or\"ok}
	\affiliation{Department of Physics, Budapest University of Technology and Economics, Budafoki ut 8, 1111 Budapest, Hungary}
	\affiliation{MTA-BME Condensed Matter Research Group, Budafoki ut 8, 1111 Budapest, Hungary}
	
	\author{Mikl\'os Csontos}
	\affiliation{Department of Physics, Budapest University of Technology and Economics, Budafoki ut 8, 1111 Budapest, Hungary}
	\affiliation{Empa, Swiss Federal Laboratories for Materials Science and Technology, Transport at Nanoscale Interfaces Laboratory, \"Uberlandstrasse 129, CH-8600 D\"ubendorf, Switzerland}
	
	\author{P\'eter Makk}
	\affiliation{Department of Physics, Budapest University of Technology and Economics, Budafoki ut 8, 1111 Budapest, Hungary} 
	
	\author{Andr\'as Halbritter$^{*}$}
	\email{halbritt@mail.bme.hu}
	\affiliation{Department of Physics, Budapest University of Technology and Economics, Budafoki ut 8, 1111 Budapest, Hungary}
	\affiliation{MTA-BME Condensed Matter Research Group, Budafoki ut 8, 1111 Budapest, Hungary}
	
	\title{Breaking the quantum PIN code of atomic synapses}
	
	\begin{abstract}
		Atomic synapses represent a special class of memristors whose operation relies on the formation of metallic nanofilaments bridging two electrodes across an
		insulator. Due to the magnifying effect of this narrowest cross-section on the device conductance, a nanometer scale displacement of a few atoms grants access
		to various resistive states at ultimately low energy costs, satisfying the fundamental requirements of neuromorphic computing hardware. Yet, device engineering
		lacks the complete quantum characterization of such filamentary conductance. Here we analyze multiple Andreev reflection processes emerging at the filament
		terminals when superconducting electrodes are utilized. Thereby the quantum PIN code, i.e. the transmission probabilities of each individual conduction channel
		contributing to the conductance of the nanojunctions is revealed. Our measurements on Nb$_2$O$_5$ resistive switching junctions provide a profound experimental evidence that the onset of the high conductance ON state is
		manifested via the formation of truly atomic-sized metallic filaments.
	\end{abstract}

\date{\today}
\maketitle

Keywords: resistive switching, memristor, superconductivity, atomic junction, niobium, niobium oxide

\section{Introduction}

\begin{figure}[b!]
\centering
\includegraphics[width=0.48\textwidth]{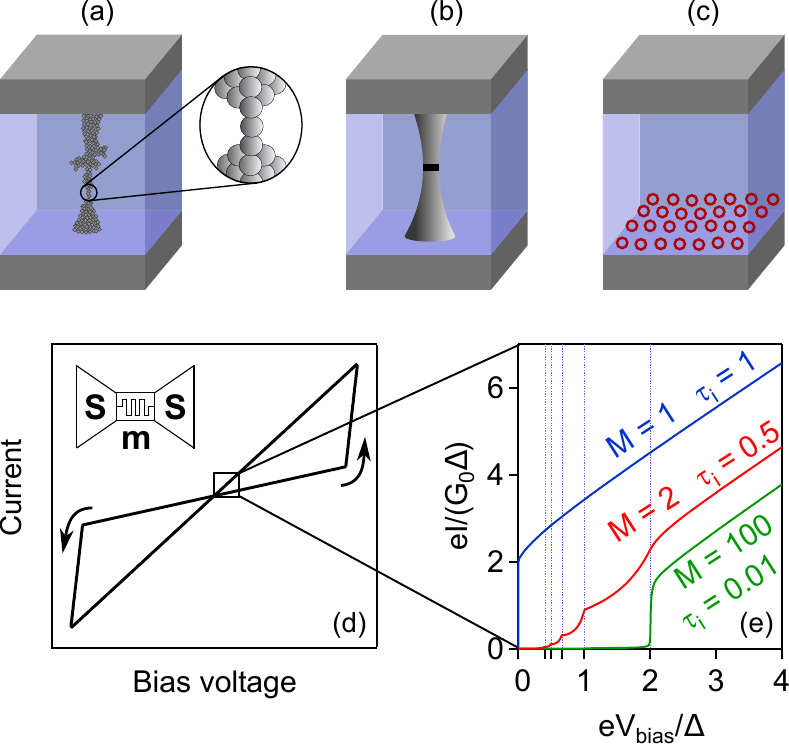}
    \caption{\it \textbf{The scheme of our analysis.} Top panels: various fundamentally different resistive switching memory arrangements: (a) truly single-atom
    diameter filamentary switch; (b) filamentary switch with a few nm filament diameter and a thin tunnel barrier at the middle; (c) interface-type device where
    the filling/emptying of oxygen vacancies at the interface (red dots) is responsible for the switching~\cite{Sawa2008}, and the current flows though a large
    cross-sectional area tunnel junction. When such memristive devices are terminated by superconducting electrodes (see the inset in panel (d)) the distinct structures
    in the superconducting subgap characteristics (e) provide a unique possibility to distinguish different types of junctions even if their switching $I(V)$
    characteristics (d) would share device states with the same, $\approx 1$\,G$_0$ conductance. Panels (a-d) are illustrations, whereas panel (e) demonstrates
    distinct theoretical subgap $I(V)$ curves~\cite{Rubio-Bollinger-subgap} for junctions with the same $1$\,G$_0$ conductance but different $\tau_i$ and $M$ values.}
    \label{fig1}
\end{figure}

Recently an incredible progress has been achieved in the hardware
implementation of artificial neural networks utilizing resistive switching memory
(RRAM) technology relying on the voltage induced formation and degradation of conducting filaments within an insulator matrix~\cite{Chua1971,Strukov2008a,Chung2010,Yang2013a, Zidan2018,Xia2019}.
As an example, 128x64 memristor crossbar arrays were built and successfully applied for efficient image processing and machine learning tasks~\cite{Li2018,Li2019,Li2018NatComm,Hu2018}. Such \emph {artificial synapse} devices usually exploit the highly linear current-voltage characteristics and the broad analog tunability of the resistance states in their transition metal oxide memristor units, which are typically operated in the
$<1000\,\mu$S conductance range approaching or
spanning the $G_{0}=2e^{2}/h\approx77.5$\,$\mu$S universal conductance quantum~\cite{Prezioso2014,Li2018,Jiang2018,Li2019,Wang2019,Li2018NatComm,Hu2018,Xia2019}. In the latter
regime it is tempting to interpret the RRAM device state as an atomic-sized filament (Fig.~\ref{fig1}a), representing the ultimate smallest memory element. However, it is evident that solely
the conductance value cannot supply any information about the cross sectional area of the
active device region: a truly atomic-sized metallic filament may provide exactly the same
conductance as a much wider, nanometer-scale filamentary switch with a tunnel junction
at the middle (Fig.~\ref{fig1}b), or an even larger interface-type RRAM device (Fig.~\ref{fig1}c)~\cite{Sawa2008}.

At truly atomic dimensions the direct microscopic imaging of the active volume of
resistive switching devices is extremely challenging. Therefore, the claims on atomic scale
switching typically rely on indirect evidences. For instance, elemental single-atom silver
nanowires are known to exhibit a well-defined configuration with a conductance of $2e^{2}/h$,
therefore the statistically pronounced occurrence of the quantum conductance in silver-based filamentary RRAM units is an indication of atomic switching~\cite{Aono2010,Wagenaar2012,Cheng2019}. On the contrary, pure single-atom nanowires made of transition
metal elements exhibit very broad conductance distributions, where the conductance quanta are not distinguished in any sense~\cite{Ruitenbeek-PhysRep-quantum,Makk2008}. Moreover, the variable oxygen content of the conducting filaments in
transition metal oxide-based RRAMs is expected to further increase the conductance
variety. As a consequence, it is extremely challenging to identify the physical nature of
the conducting filaments in these technologically highly important structures.

Here we employ the powerful method of superconducting subgap spectroscopy
developed in the field of mesoscopic physics~\cite{Scheer1998,Ruitenbeek-PhysRep-quantum,Ludolph2000,Makk2008,res-switch-book,Cuevas2010}. This method is capable of decomposing all
the $\tau_i$ transmission probabilities (the so-called quantum PIN code~\cite{mesoscopic-PIN,Ruitenbeek-PhysRep-quantum}) of the individual
quantum conductance channels contributing to filamentary conductance, thus provides
substantially more information about the conduction properties than the overall
$G=\frac{2e^2}{h}\sum_{i=1}^{M}\tau_i$ conductance~\cite{Landauer1970}, where $M$ is the number of
open quantum conductance channels. This approach was originally implemented in the field of
atomic and molecular electronics to reveal the nature of conductance in single atom
nanowires~\cite{Scheer1998} and more recently to identify the
distinct atomic states upon reversible current-induced single-atom rearrangements~\cite{Scheer2013}. Here we apply this unique method
to study the nature of the conducting filaments in transition metal oxide
based RRAM structures. We focused our studies on Nb/Nb$_{2}$O$_{5}$/Nb point contacts where the advantageous resistive switching properties~\cite{Gong2018,Wang2019,Jiang2018,Li2018,Li2019,Torrezan2011,Strukov2008a,Prezioso2014,Pickett2012,Mahne2014,Nb2O5-neuromorphic2018,Liu2012b,Wylezich2015,Wylezich2014,Kim2012b} are accompanied by the conveniently high superconducting transition temperature ($T_{c}$=9.22~K) of the elemental Nb electrodes.
Our measurements provide a profound experimental evidence that the observed switching takes place due to
the structural rearrangement of a truly single-atom diameter conductance channel.

The scheme of our analysis is illustrated in Fig.~\ref{fig1}d,e. At higher voltage scales the resistive switching junction exhibits a conventional hysteretic $I(V)$
characteristics (Fig.~\ref{fig1}d). However, if the $I(V)$ traces are compared in the range of the superconducting gap ($\Delta$), distinct structures are
observed due to multiple Andreev reflections (Fig.~\ref{fig1}e)\cite{Ludolph2000,Scheer1998,Cuevas2010,Rubio-Bollinger-subgap}. As a first order process, single electron
charges can pass the junction with $\tau$ probability, but due to the presence of the superconducting gap, this is only possible at $eV>2\Delta$. However, an
$n^{\textrm{th}}$ order process including the simultaneous transfer of $n$ electron charges with $\tau^{n}$ probability becomes available at a reduced voltage
of $eV>2\Delta/n$. In a tunnel junction all the transmission probabilities are small ($\tau_i\ll1$) and therefore all the higher order processes are negligible. In this case the current remains zero at $eV<2\Delta$,
whereas at higher voltage a linear $I(V)$ curve is observed with the slope of the $G_{N}$ normal state conductance (see the green curve in Fig.~\ref{fig1}e).
In an atomic-sized metallic filament, however, a single or a few conductance channels are highly transparent (i.e. their transmission probability is close to unity), and so the higher order processes also become enabled. This introduces finite \emph{subgap} current at $eV<2\Delta$
with distinct structures at the $2\Delta/n$ thresholds (red curve in Fig.~\ref{fig1}e). In the extreme case of $\tau_i=1$ even the $n \gg 1$ order
processes are available giving rise to an infinitely steep current rise at zero voltage (blue curve in Fig.~\ref{fig1}e). By the numerical fitting of the $I(V)$
curve in the gap region one can, in principle, determine all the $\tau_i$ transmission eigenvalues\cite{Rubio-Bollinger-subgap}, and thus one can clearly distinguish physically different device states even if they share the same conductance. 

\section{Results and Discussion}
Before demonstrating our main result of resolving truly atomic-scale resistive switching by subgap spectroscopy, we take the following steps: (i) we demonstrate the operation of Nb$_2$O$_5$ resistive switching junctions close to the quantum conductance unit also highlighting the analog tunability of the resistance states in this regime; (ii) we present reference measurements on pure Nb atomic junctions also demonstrating the proper spectroscopic resolution of our subgap spectroscopy setup; (iii) we analyze how much the applicability of subgap spectroscopy is restricted by the superconducting proximity effect in the niobium oxide region.

\begin{figure}[t!]
\centering
\includegraphics[width=0.38\textwidth]{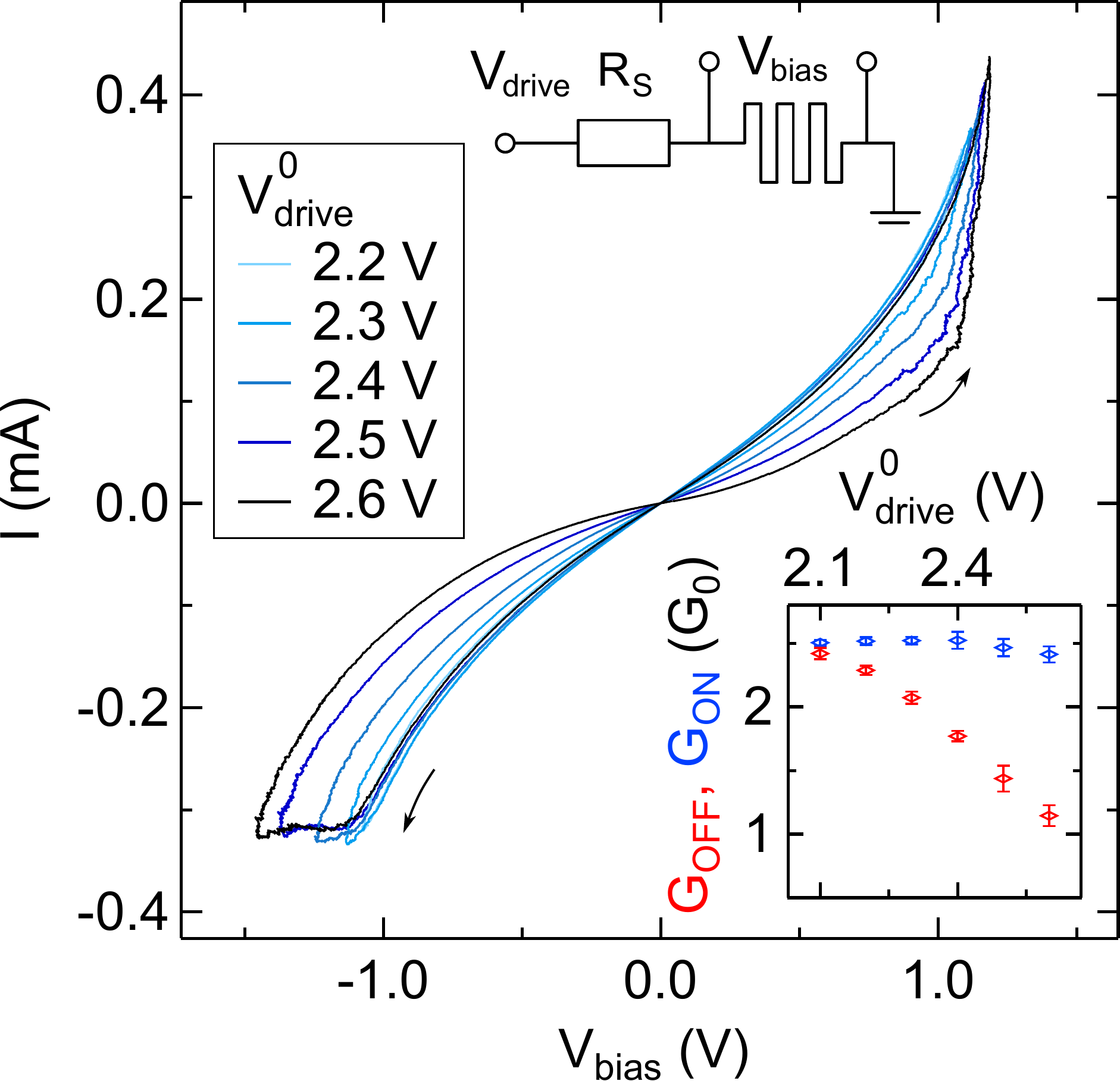}
    \caption{\it \textbf{Multilevel programming in Nb$_2$O$_5$ resistive switching junctions set to the vicinity of the quantum conductance unit.} The resistive switching junction and the $R_S$ serial resistor are driven by a $V_{\rm drive}$ triangular signal (top inset), and the $I(V)$ characteristics are displayed as a function of the voltage drop on the memristor junction, $V_{\rm bias}=V_{\rm drive}-R_S\cdot I$. As the $V_{\rm drive}^0$ amplitude is increased the hysteretic $I(V)$ curves are opening up. The bottom inset exhibits the $V_{\rm drive}^0$ dependence of the $G_{\rm ON}$ and $G_{\rm OFF}$ low voltage conductances, i.e. the slopes of the $I(V)$ curves in the $\pm100\,$mV region. The error bars represent the standard deviations calculated from 8 consecutive $I(V)$ curves with fixed $V_{\rm drive}^0$.}
    \label{fig2}
\end{figure}

\subparagraph*{Resistive switching in the vicinity of the quantum conductance unit}  
Our measurements were performed on $\approx20$\,nm thick Nb$_2$O$_5$ layers that were grown on the top of a $\approx300$\,nm thick Nb thin films by anodic oxidation. The resistive switching junctions were established in a scanning tunneling microscope (STM) arrangement by touching the PtIr STM tip to the thin film sample.
The sample preparation and the scheme of the measurement follows the same protocol as in our previous study on the general resistive switching properties of Nb$_2$O$_5$\cite{Nb2O5-neuromorphic2018}. Here we demonstrate, that Nb$_2$O$_5$ exhibits room temperature resistive switching in the vicinity of the quantum conductance unit as well (Fig.~\ref{fig2}).
Furthermore, as the $V_\textrm{drive}^0$ amplitude of the driving triangular signal increases, the resistive switching curves open up exhibiting a clear multilevel programmability. Accordingly, the device states can be fine tuned in the $\approx 1\,$G$_0-2.5\,$G$_0$ interval as demonstrated by the $V_{\rm drive}^0$ dependence of the $G_{\rm ON}$ and $G_{\rm OFF}$ low voltage conductances in the bottom inset of Fig.~\ref{fig2}. We emphasize that in spite of the $\approx 1\,$G$_0$ quantum conductance range, no conductance jumps due to distinct atomic rearrangements are observed, the OFF state conductance is rather tunable in a fully continuous fashion. In this case the ON state conductance remains constant, which we attribute to the interplay of the $R_S=3.35\,$k$\Omega$ serial resistance and the strong intrinsic nonlinearity of the $I(V)$ curve\cite{Nb2O5-neuromorphic2018} restricting the $V_{\rm bias}$ voltage drop on the junction.

To study the superconducting subgap characteristics of such resistive switching junctions we have performed our further measurements in an STM setup operated at $T=1.4$\,K temperature. This low temperature setup is optimized to prevent noise pickups, which would induce a smearing of the spectroscopic information in the subgap $I(V)$ curves. 
Further details on the sample preparation, the measurement protocol and the electronic circuitry including the various filter stages are provided in the Methods
section.

\subparagraph*{Reference experiments on pure Nb atomic wires}

\begin{figure}[t!]
 \centering
 \includegraphics[width=0.48\textwidth]{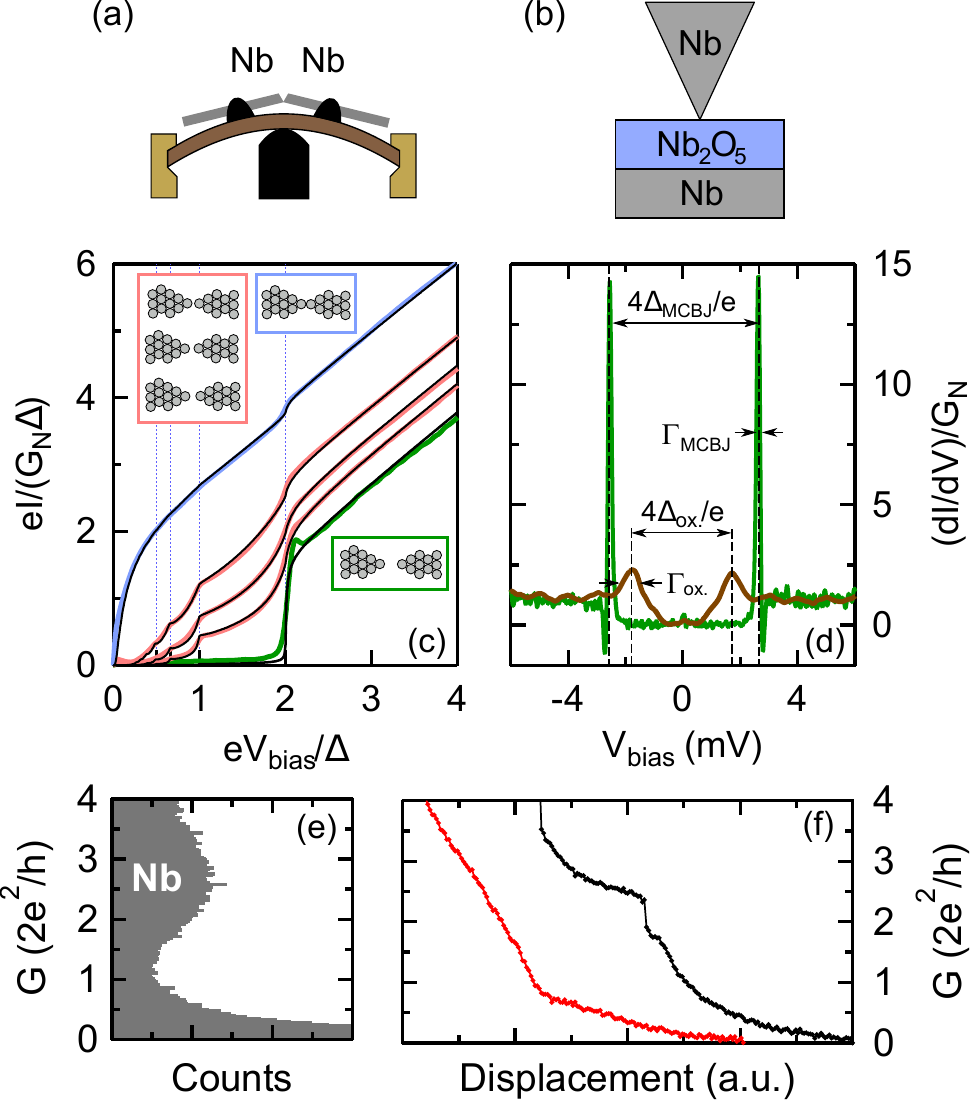}
    \caption{\it \textbf{Reference measurements on clean Nb atomic contacts.} Illustration of the MCBJ (a) and STM (b) arrangements (see Methods for more details).
    (c) Representative $I(V)$ characteristics measured on pure Nb single-atom junctions. The transmission eigenvalues and the normal state conductances obtained
    from the fits (black curves) going from the top to the bottom curve are
    $\tau_i$=\{0.969, 0.112, 0.022, 0.021, 0.020\},\{0.686, 0.142, 0.017, 0.014, 0.013\}, \{0.514, 0.148, 0.024, 0, 0\}, \{0.344, 0.032, 0.032, 0, 0\}, \{0.016, 0, 0, 0, 0\};
    $G_N/G_0=\sum_i\tau_i$=1.144, 0.872, 0.686, 0.408, 0.016. All curves were fitted using five conductance channels. The insets illustrate possible atomic arrangements
    behind these subgap curves (the color of the frames refers to the corresponding curves). (d) Differential conductance of Nb/Nb (green line, $G_N=$0.016\,G$_0$)
    and Nb/Nb$_2$O$_5$/Nb (brown line, $G_N=$0.0088\,G$_0$) tunnel junctions. The values of the energy gap are $\Delta_{\rm MCBJ} = 1.294$~mV and $\Delta_{\rm ox.}=0.866$~mV,
    and the width of the characteristic peaks at $eV=\pm2\Delta$ are $\Gamma_{\rm MCBJ}=131~\mu$V and $\Gamma_{\rm ox.}=565~\mu$V, respectively.
    (e) A typical conductance histogram of Nb nanowires based on 10000 repeating breaking cycles measured with a break junction setup. (f) Sample conductance vs.\ electrode separation traces exhibiting a rather smooth variation of the conductance as the atomic-sized Nb junction is disconnected by the piezo actuator. 
    }
    \label{fig3}
\end{figure}

Prior to resistive switching measurements we have characterized our low temperature subgap spectroscopy setup
using the well-studied reference system of pure single-atom Nb nanojunctions~\cite{Ludolph2000,Makk2008,Nb-dimer-subgap,Scheer1998} established in a mechanically
controllable break junction (MCBJ) arrangement (see Fig.~\ref{fig3}a). In this case a macroscopic Nb wire is broken in a three point bending configuration to
form extremely stable single-atom contacts, which are ultra-clean due to the freshly broken surfaces. Fig.~\ref{fig3}c displays the experimental subgap curves
of pure atomic-sized Nb contacts realized at different displacements of the electrodes. For a better comparison of the subgap curves corresponding to different
$G_N$ values, the conventional normalization procedure of the current and voltage scales is applied~\cite{Ludolph2000}, such that all curves scale to a slope
of unity at $eV_\textrm{bias}/\Delta\gg2$. The bottom, green curve in Fig.~\ref{fig3}c shows a typical tunneling characteristics resembling the green curve in Fig.~\ref{fig1}e. The numerical derivative of this $I(V)$ curve shows sharp peaks at $\pm2\Delta/e$ (see the green differential conductance curve in Fig.~\ref{fig3}d). The $\Gamma_{\rm MCBJ}=131\,\mu$V half-width of these peaks directly tells the voltage resolution of our subgap measurement setup. As our MCBJ and STM setups are exact clones of each other (apart from the mechanical actuation), this voltage resolution can also be considered as an electronic resolution baseline for our resistive switching experiments. 

At such resolution all the subgap $I(V)$ curves in Fig.~\ref{fig3}c are well fitted with the theory of multiple Andreev reflections\cite{Ludolph2000,Scheer1998,Cuevas2010,Rubio-Bollinger-subgap} (see the black fitting curves in Fig.~\ref{fig3}c and the the corresponding transmission eigenvalues in the caption). Further details on the fitting procedure are provided in the Methods section. 

The transmission eigenvalue decomposition reveals that the transport is dominated by the first conductance channel for all traces, exhibiting increasing $\tau_1$ from the bottom (green) curve
towards the top (blue). The $\tau_1\ll1$ value for the green curve confirms that a tunneling junction is concerned, the red curves correspond to partially open channels ($\tau_1\approx0.3-0.7$), whereas
the blue curve resembles the blue curve in Fig.~\ref{fig1}e representing a single dominant channel with nearly perfect transmission ($\tau_1\approx0.97$). Note, that the transition
from a tunnel junction to a transparent metallic nanowire is not only indicated by the transition from zero current to a steep current rise in the subgap
regime ($eV_\textrm{bias}/\Delta<2$). At the same time the so-called {\it excess current} is also increased, i.e. the high-bias ($eV_\textrm{bias}/\Delta\gg2$)
linearly varying part of the curves exhibits an increasing current offset as the channels open up.

Next, we briefly review the well-studied transmission properties of Nb single-atom nanowires\cite{Ludolph2000,Scheer1998,Makk2008,Ruitenbeek-PhysRep-quantum}, which will serve as a comparison basis for our subgap analysis on Nb$_2$O$_5$ resistive switching junctions. In a simple free electron picture one can argue that the first quantum conductance channel opens in nanowires, where the $(2\pi\hbar)^2/ (2\lambda_F^2m^*)$
kinetic energy of the electrons at the Fermi surface of the electrodes exceeds the transverse confinement energy at the narrowest cross section of the wire. 
Considering a cylindrical nanowire geometry \cite{Ruitenbeek-PhysRep-quantum} and the $\lambda_F\approx0.53\,$nm Fermi wavelength \cite{Ashkroft-Mermin} of niobium the first quantum conductance channel is expected to open at $R\approx0.2$\,nm filament radius, i.e. the first channel indeed opens at truly atomic dimensions. However, it is to be emphasized that the free electron picture is a very rough approximation in transition metal nanowires \cite{Ruitenbeek-PhysRep-quantum}, more realistic first principle
simulations and subgap spectroscopy measurements refine this picture showing that a single atom diameter Nb nanowire has a broad conductance
distribution around $2.5\,G_0$ possessing up to five partially open channels due to the transport through the $s$ and $d$ valence orbitals of the central atoms~\cite{Scheer1998}. The transport through the $d$ orbitals happens through partially open channels, for which $\tau_i$ are mostly well off from unity, whereas the $s$ channel is usually well transmitting~\cite{Cuevas1998}. Furthermore, the transport through the $d$ channels is very sensitive to the precise details of the particular atomic arrangement. As a clear consequence, one should not expect any sign of conductance quantization features, rather a broad continuum of possible conductance values appears. To illustrate this we reproduce a typical conductance histogram of Nb in Fig.~\ref{fig3}e (See Refs.~\citenum{Ruitenbeek-PhysRep-quantum,Ludolph2000,Makk2008}) demonstrating  that \emph{any conductance value} can be set in the plotted $G=0-4\,$G$_0$ range, and the quantized values are not enhanced at all. The sample conductance versus electrode separation traces in Fig.~\ref{fig3}f also illustrate that in Nb (and in various further transition metals) the well-known conductance staircase of noble metal nanowires \cite{Ruitenbeek-PhysRep-quantum} is replaced by a rather smooth and continuous conductance variation with minor conductance jumps (black curve) or no conductance jumps at all (red curve). 

According to the above considerations the atomic configurations behind the subgap curves of Fig.~\ref{fig3}c are reflecting the smooth disconnection of a single-atom nanowire along the continuous $G\lesssim 1\,$G$_0$ tail region of the conductance traces in Fig.~\ref{fig3}e. This is illustrated with the inset cartoons in Fig.~\ref{fig3}c:
the blue curve corresponds to a transparent single atom junction, the red curves are related to junctions, where the central atoms are already slightly
disconnected, and the green curve reflects a disconnected tunneling junction.

\subparagraph*{Preconditions of subgap spectroscopy on resistive switching junctions.} The application of superconducting subgap spectroscopy on the ON and OFF
states of resistive switching Nb/Nb$_2$O$_5$/Nb junctions relies on three obvious preconditions: (i) resistive switching should work with a compositionally
symmetric electrode arrangement, i.e. using Nb electrodes on both sides; (ii) operation at cryogenic temperatures; (iii) the presence of the oxide layer should
not result in an untolerable reduction of the subgap spectroscopy's resolution. In the following these requirements are analyzed.

\begin{figure}[t!]
	\centering
	\includegraphics[width=0.48\textwidth]{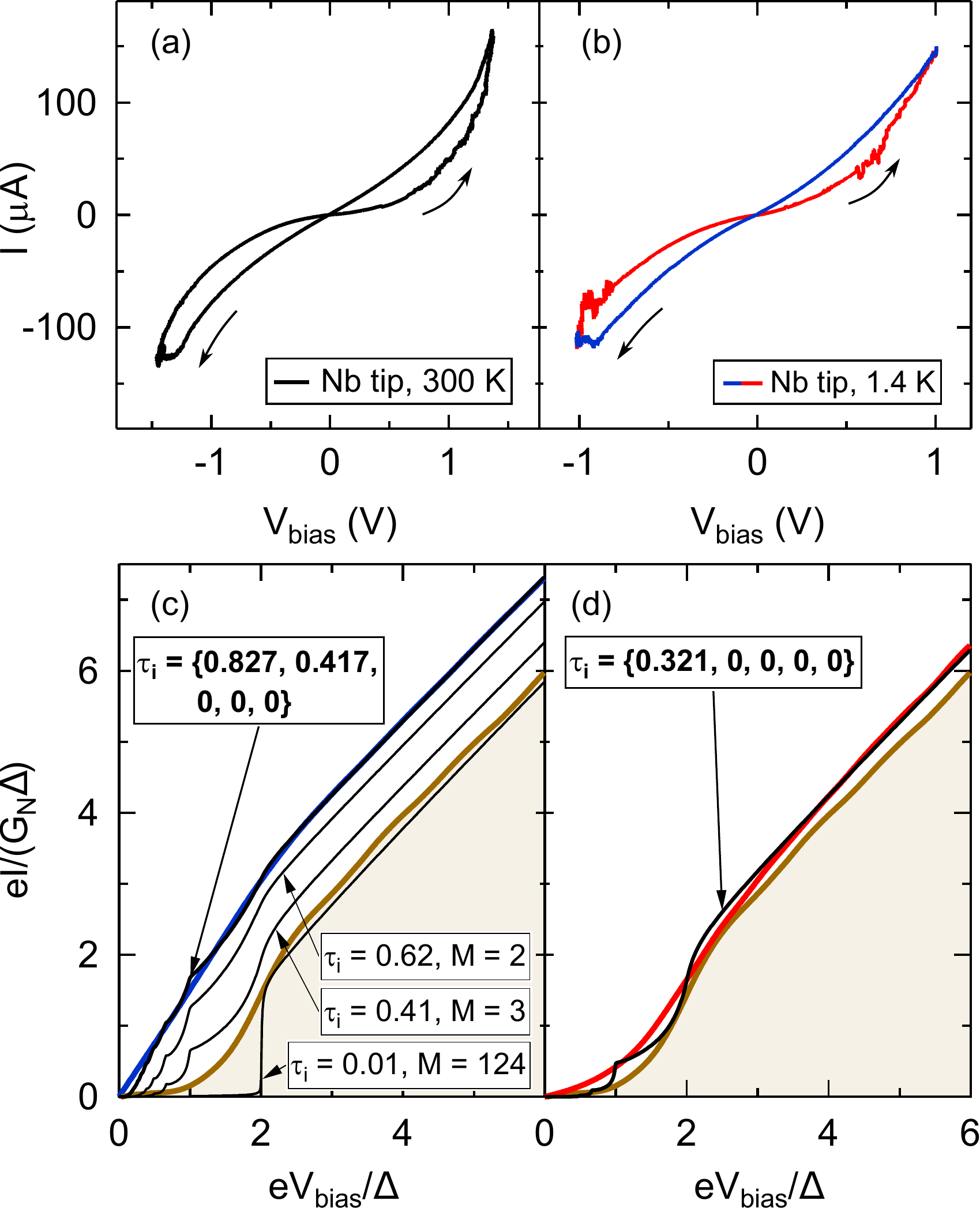}
	\caption{\it \textbf{Quantum PIN code decomposition of the ON and OFF resistance states.} (a) Representative resistive switching characteristic acquired in
		 a Nb(tip)/Nb$_2$O$_5$/Nb(thin film) junction at room temperature. (b) Resistive switching
		observed in a Nb(tip)/Nb$_2$O$_5$/Nb(thin film) junction at $T=1.4\,$K. The red (blue) part corresponds to the OFF (ON) state, the arrows illustrate the direction
		of the hysteresis. The blue and red curves in panels (c) and (d) show the subgap $I(V)$ traces measured in the ON and OFF states of the corresponding
		resistive switching characteristics of panel (b), respectively. The thick black lines demonstrate the best fitting theoretical $I(V)$ curves using five open conductance
		channels (see the corresponding boxes for the fitted transmission eigenvalues). As a comparison, the thin black curves in panel (c) illustrate theoretical subgap traces
		where the total conductances identical to the conductance of the measured ON state are shared between different numbers of equally transmitting channels
		(see the corresponding boxes). As a reference, the brown lines show a scaled tunneling characteristics with $G_N\approx0.01\,$G$_0$ conductance measured
		on a disconnected junction.}
	\label{fig4}
\end{figure}

(i) Most works apply a compositionally asymmetric junction design to grant a well-defined bipolar resistive switching, however, in our case subgap spectroscopy necessitates Nb electrodes on both sides. In our previous work \cite{Gubicza2016} we have demonstrated that the tip-sample geometrical asymmetry alone is enough to enable bipolar resistive switching. This was justified on our specific Nb$_2$O$_5$ resistive switching junctions as well, as demonstrated by the similar room temperature switching
$I(V)$ traces using PtIr(tip)/Nb$_2$O$_5$/Nb(thin film) junctions (Fig.~\ref{fig2}) or Nb(tip)/Nb$_2$O$_5$/Nb(thin film) junctions (Fig.~\ref{fig4}a). Based on the statistical analysis of 100 independent junctions we generally find, that in spite of the symmetric electrode material
arrangement the Nb(tip)/Nb$_2$O$_5$/Nb(thin film) junctions exhibit a dominant switching voltage polarity: in $>80\%$ of the cases the set transition happens
when the sample is positively biased with respect to the tip. In the remaining cases the local geometrical asymmetry of the filament center is presumably reversed compared to the larger scale tip-sample asymmetry.

(ii) Our low temperature measurements ($T=1.4$\,K) have routinely yielded resistive switching curves, which are similar to the room temperature switching
characteristics (see a typical low temperature switching curve of Nb/Nb$_2$O$_5$/Nb junctions in Fig.~\ref{fig4}b). The low temperature operation of the
switching is attributed to the extremely large electric fields at the narrowest part of the junction as well as to the self-heating effect of the active
junction area~\cite{Gubicza2015b}.

(iii) The superconducting features are clearly observed in the $I(V)$ curves of the SmS junctions. However, according to the differential conductance curve of
a Nb/Nb$_2$O$_5$/Nb tunnel junction (brown curve in Fig.~\ref{fig3}d) the width of the superconducting coherence peak is increased ($\Gamma_{\rm ox.}=565\,\mu$V), whereas the gap value determined
from the peak position ($\Delta_{\rm ox.}=0.866$~mV) is reduced with respect to the clean Nb MCBJ junctions. This $\Gamma$-broadening results in a smearing of all subgap traces, and so the spectroscopic resolution is reduced. In the following we discuss the possible
background of the resolution loss, emphasizing that our subgap data are still suitable to draw the conclusions of our study. As the STM and the
MCBJ setups share the same electromagnetic environment thank to the same sample holder structure and measurement circuits, including identical filter stages,
we exclude the possibility of enhanced noise pickups in the former case. According to the XPS analysis carried out in our earlier
study,~\cite{Nb2O5-neuromorphic2018} our Nb$_{2}$O$_{5}$/Nb thin film samples contain an $\approx10\,$nm thick interface region of inhomogeneous oxygen content
between the Nb$_2$O$_5$ layer and the bulk Nb. We argue that this suboxide region forms a conducting, but intrinsically non-superconducting
volume~\cite{Halbritter-Nb,ScThin-Scirep}, which is made superconducting by the proximity effect of the nearby superconducting Nb electrode. Such proximity
superconducting structures are known to exhibit a reduced gap value and a smeared superconducting density of states~\cite{Moussy2011,Vinet2000,Shen1972,leSueur2008}, as it
was also demonstrated along the subgap spectroscopy of Al/Au/Al atomic contacts~\cite{Proximity-Scheer}. As a rough estimate based on the theoretical model
described in Refs.~\citenum{Belzig1996,Proximity-Scheer}, the presence of a $10\,$nm wide proximity superconducting region would induce the observed
$\Delta_{\rm ox.}/\Delta_{\rm MCBJ} \approx 0.67$ reduction of the measured gap value in our oxide samples (see Fig.~\ref{fig3}d) if a superconducting
coherence length of $\approx22\,$nm is assumed. The latter value is reasonable in a highly disordered oxide layer\cite{Forro-NbO} in comparison with the $39\,$nm bulk
coherence length of niobium.~\cite{Kittel2005}

\subparagraph*{Quantum PIN code decomposition of the ON and OFF resistance states.} Having the basic requirements of subgap spectroscopy satisfied, we wish to
classify our resistive switching junctions via their quantummechanical PIN code decomposition. If a larger area tunneling junction (Fig.~\ref{fig1}b,c) is
concerned, the green tunneling characteristic of Fig.~\ref{fig1}e should be measured. However, the $\Gamma$-broadening yields a smearing of this curve,
as demonstrated by the brown lines in Fig.~\ref{fig4}c,d showing an experimentally measured tunneling trace with $G_N\approx0.01\,$G$_0$ conductance. Note,
that if the $\tau_i \ll 1$ condition is satisfied, the tunneling $I(V)$ curves scale to the same universal dimensionless trace on the $eI/(G_N\Delta)$ vs.
$eV_\textrm{bias}/\Delta$ plane of Fig.~\ref{fig4}c,d. This means that the brown curves in panels (c) and (d) are expected to look similar for any tunnel
junction with arbitrary conductance. Due to the $\Gamma$-broadening, the light brown area under these brown tunneling characteristics is experimentally
unaccessible, however, any subgap trace growing above this brown background should be related to a device state which is definitely not a tunneling junction.

\begin{figure}[t!]
 \centering
\includegraphics[width=0.48\textwidth]{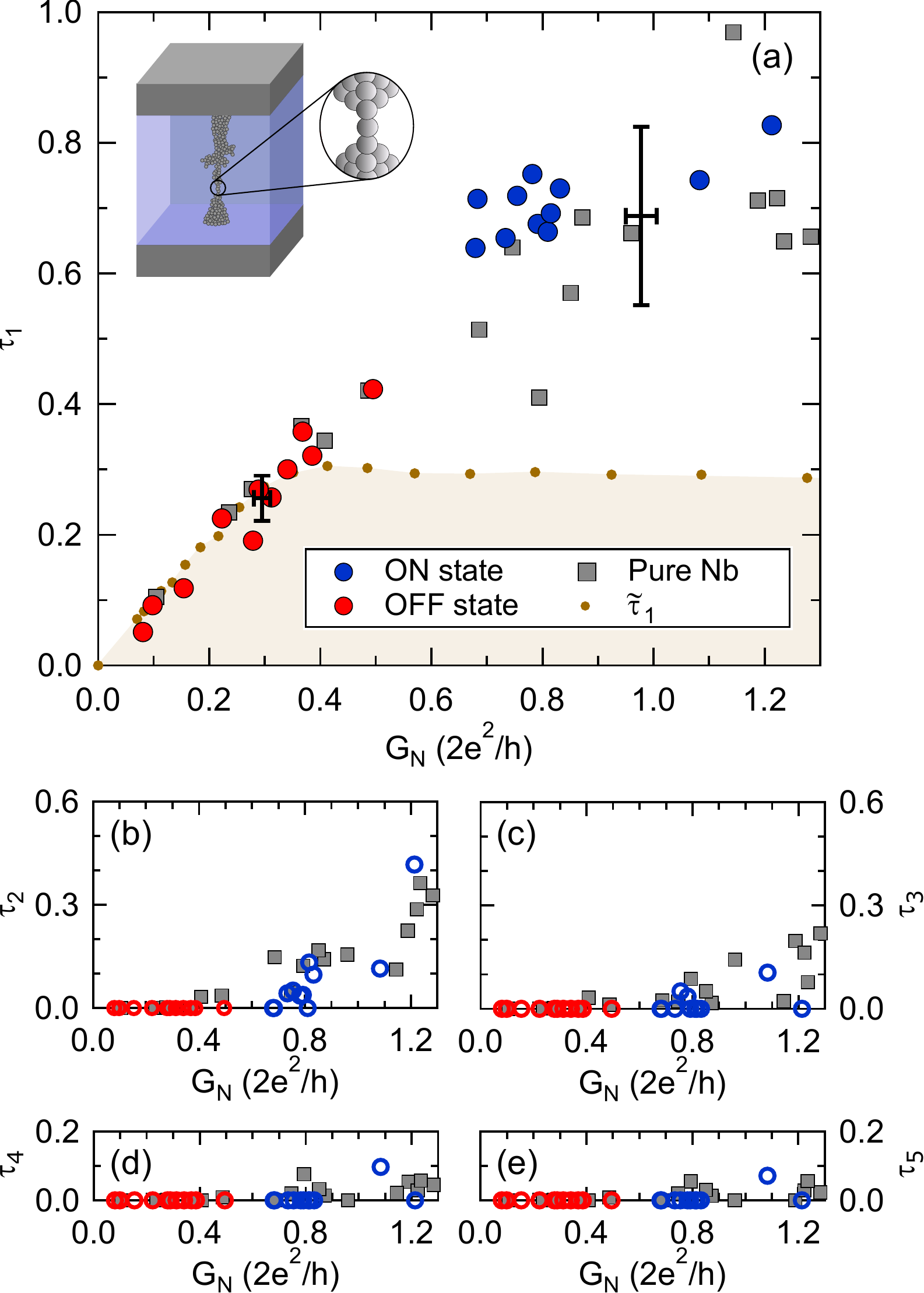}
    \caption{\it \textbf{Evolution of the conductance channels.} (a,b,c,d,e) The distribution of the $\tau_1,\tau_2,\tau_3,\tau_4,\tau_5$ transmission eigenvalues
    numerically evaluated in various independent Nb(tip)/Nb$_2$O$_5$/Nb(thin film) resistive switching junctions measured at $T=1.4\,$K. The blue (red) circles
    correspond to the ON (OFF) states, respectively. As a comparison, the grey squares display the transmission eigenvalues of pure Nb single-atom nanowires also
    including the data shown in Fig.~\ref{fig3}c. As a further reference taken from Ref.~\citenum{Makk2008}, the black datapoints and error bars represent the
    mean transmission values and their standard deviations acquired on 30$+$30 independent pure Nb atomic junctions measured at $G\approx1\,$G$_0$ and
    $G\approx0.3\,$G$_0$ conductances. The
    shaded area in panel (a) highlights the region in the $\tau_1 - G_N$ plane where our analysis is not conclusive due to the $\Gamma$-broadening (see text).
    The inset in panel (a) illustrates an atomic-sized filament.}
    \label{fig5}
\end{figure}

The blue and red lines in Fig.~\ref{fig4}c,d respectively show the subgap $I(V)$ curves of the ON and OFF states demonstrated in Fig.~\ref{fig4}b. 
The subgap curve of the ON state with $G_N=1.244$\,G$_0$ conductance (Fig.~\ref{fig4}c) clearly separates from the brown background exhibiting a steep current rise around
zero bias similarly to the blue curves in Fig.~\ref{fig1}e and Fig.~\ref{fig3}c. For this curve the fitting procedure unambiguously concludes a dominant channel with close to unity transmission, which is extended by a further channel with smaller transmission (see the thick black fitting curve and the corresponding PIN code). This result provides a clear evidence that the ON state corresponds to a highly transmitting filament with a single atom
at the narrowest cross section (see the illustration in Fig.~\ref{fig1}a). If the same conductance would be shared between a larger number of less transmitting
channels, the $I(V)$ curve would strongly deviate from the measured curve, as illustrated by the thin black theoretical subgap traces owing the total
conductance of the ON state shared between different number of equally transmitting channels. On the other hand, a highly transmitting filament with
several atoms in the narrowest cross section would correspond to a filament diameter significantly exceeding the $\lambda_F=0.53\,$nm Fermi-wavelength, and
therefore it is expected to exhibit a larger conductance with more than one highly transmitting channels~\cite{Ruitenbeek-PhysRep-quantum,res-switch-book}.
The subgap trace of the OFF state (Fig.~\ref{fig4}d) is best fitted with a single conductance channel, as demonstrated by the thick black line and the corresponding set of transmission eigenvalues. However, this subgap curve only slightly grows above the brown background, indicating that
the $G_N=0.321$\,G$_0$ conductance of the OFF state is already close to the border, where subgap spectroscopy provides a less conclusive classification between a single-channel conductor or a multi-channel tunnel junction. 
In spite of this uncertainty, we argue that a single-atom diameter ON state is expected to switch to a single-atom diameter OFF state such that a narrow barrier forms between the central atoms due to a voltage induced atomic displacement at the junction center.

\subparagraph*{Statistical analysis of the transmission eigenvalues.} Next, we investigate the transmission properties of a larger ensemble of
Nb/Nb$_2$O$_5$/Nb resistive switching junctions (see Fig.~\ref{fig5}). To precisely define the validity range of our analysis, we linearly re-scale the brown tunneling
curve of Fig.~\ref{fig4}c,d to various normal state conductances, $\tilde{G}_N$. We fit these scaled tunneling curves and extract $\tilde{\tau}_1$, the leading transmission
probability values (brown dots in Fig.~\ref{fig5}a). Due to the $G_N\approx0.01\,$G$_0$ conductance and the corresponding $\tau_1\le0.01$ transmission
probability of the original tunneling curve, the re-scaled curves should exhibit similarly small $\tau_1$ values with $M\ge \tilde{G}_N/G_N$ conductance channels. The fitting procedure, however, provides only a few channels with 
significantly larger $\tilde{\tau}_1$ values due to the smearing of the tunneling curves.
As a general tendency, at $\tilde{G}_N < 0.3$\,G$_0$ the fitting yields a single channel with $\tilde{\tau}_1\approx\tilde{G}_N/$\,G$_0$, whereas at $\tilde{G}_N>0.3$\,G$_0$, the leading transmission saturates at $\tilde{\tau}_1\approx 0.3$, and the remaining conductance is filled with further channels.
Accordingly, the light brown area under these
$\tilde{\tau}_1$ values defines the range, where our analysis is not conclusive. Again, the leading transmissions of the OFF states (red circles) are close to the
validity border, however, the $\tau_1$ values of the ON states (blue circles) are all well above the light brown area with $\tau_1\approx0.6-0.9$, indicating that all the investigated resistive switching junctions exhibit a single-atom wide filamentary ON state (see the inset in Fig.~\ref{fig5}a).

Finally, we compare the transmission properties of atomic-sized niobium-oxide resistive switching filaments and pure niobium atomic wires. Our previous study has already demonstrated the mean transmission probabilities and their variances at $0.3\,$G$_0$ and $1\,$G$_0$ conductances \cite{Makk2008} (black datapoints with errorbars in Fig.~\ref{fig5}a). Here we extend these data with the evolution of pure Nb transmission probabilities covering the whole conductance range, where the resistive switching was analyzed (see grey squares in Fig.~\ref{fig5}). 
The presence of oxygen in the filament may alter the transmission eigenvalues of clean Nb atomic wires in either
directions: (i) it may induce a barrier at the narrowest cross section (see the illustration in Fig.~\ref{fig1}b) yielding reduced $\tau_i$ values for the
channels with higher transmission. This effect would be especially remarkable in the reduction of $\tau_1$; (ii)  Similarly to oxygen decorated Ni atomic
wires,~\cite{Vardimon2015} the presence of oxygen may block the transport through the $d$ orbitals and enhance the role of the $s$ channel, which would result
in an increased $\tau_1$ value accompanied by the suppression of the further transmission eigenvalues. In spite of these two possibilities the data show that
the evolution of all transmission probabilities with the total conductance is very similar for resistive switching filaments and pure Nb atomic wires. This
again underlines that the transport in the $\approx1\,$G$_0$ ON state of the resistive switching niobium-oxide filaments highly resembles the transport through
single-atom Nb nanowires.

\section{Conclusions}

Concluding our analysis, we have investigated resistive switching junctions operated close to the universal quantum conductance unit. This conductance regime offers a unique possibility to establish truly atomic-sized memory devices. However, in transition metal oxide based resistive switching filaments the actual determination of the junction diameter is an especially challenging task, as the analog tunability of the conductance states is enabled even at atomic dimensions instead of displaying discrete conductance steps and quantized conductance features characteristic to noble metal atomic wires. 

Here, we have shown that superconducting subgap spectroscopy is a powerful method to gain direct insight to the
transmission properties of resistive switching junctions. Close to the quantum conductance this method is especially sensitive to the fine details of the
junction's quantum PIN code, providing highly conclusive information about the nature of the conducting filaments. Our measurements on Nb$_2$O$_5$ memristor junctions provide the first direct and well-founded experimental evidence that the switching takes place due to the structural rearrangement of a truly single-atom diameter conductance channel in a transition metal oxide resistive switching device. The method of transmission
channel decomposition can be extended to further resistive switching devices including those composed of superconducting metals (Nb, Ta, V, etc), or even
further compounds contacted with auxiliary superconducting electrodes. Furthermore, subgap spectroscopy is also adaptable for crossbar junctions utilizing superconducting electrodes, once a thin enough switching region is fabricated compared to the superconducting coherence length.   

\section*{Methods}

\subparagraph*{Preparation of atomic Nb junctions.}

Prior to the study of memristive junctions, reference subgap measurements were carried out by establishing pure atomic Nb break junctions at cryogenic
temperatures. For this purpose, 99.99~\% purity Nb wires of 0.25\,mm diameter were notched with a sharp razor in a preliminary step, followed by the insertion
of the wire into a three point bending MCBJ arrangement (see Fig.~\ref{fig3}a). A combined use of a stepper motor and a piezoelectric actuator allows broad
range actuation and precise control over breaking the wire, thus various atomic configurations (see Fig.~\ref{fig3}c inset) with stable normal conductance were
routinely achieved.

\subparagraph*{Preparation and characterization of Nb$_2$O$_5$/Nb thin films.}

Studying STM point contacts is a powerful experimental tool to characterize and optimize memristive materials in order to achieve reliable operation of future
on-chip RRAM devices. For the study of Nb(tip)/Nb$_2$O$_5$/Nb(thin film) point-contacts, the Nb$_2$O$_5$/Nb(thin film) samples were created with anodic
oxidation of a Nb thin film in a 1~\% aqueous solution of H$_3$PO$_4$, maintaining 1\,mA/cm$^2$ current density throughout the process. First a 300\,nm thick
Nb thin film was sputtered on the top of a standard Si wafer. X-ray photoelectron spectroscopy (XPS) with subsequent Ar$^+$ milling steps were performed on
Nb$_2$O$_5$/Nb(thin film) samples, uncovering the depth profile of the Nb:O stoichiometric composition. The presence of Nb$_2$O$_5$ was confirmed at the top of
the $\approx20$\,nm thick oxide layer. For further details on the anodic oxidation and the structural characterization of Nb$_2$O$_5$/Nb(thin film) samples see
Ref.~\citenum{Nb2O5-neuromorphic2018}.

\subparagraph*{Electric circuitry for resistive switching and subgap measurements.}

\begin{figure}[t!]
 \centering
\includegraphics[width=0.48\textwidth]{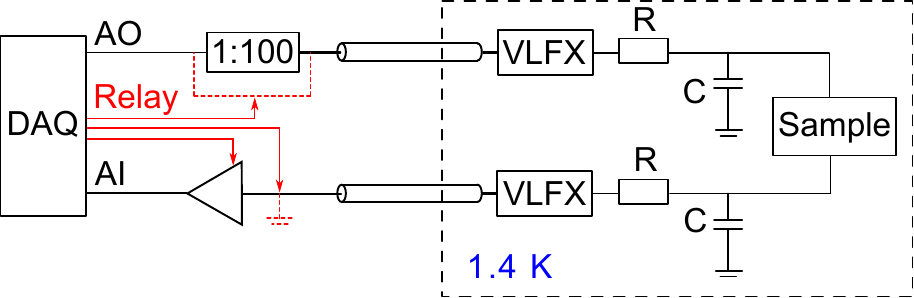}
    \caption{\it \textbf{Electric circuit diagram of the subgap measurement setup}, showing the main filtering elements: LakeShore SS-CC-100 coaxial cables
    (operating as an RC-filter), MiniCircuits VLFX-80 low pass filters, and custom-built RC-filters. The red parts refer to the automated control used for
    switching between the high-bias $I(V)$ data acquisition and subgap spectroscopy modes via two relays. The analog outputs (AO) and inputs (AI) of a
    National Instruments USB-6363 data acquisition card were utilized to bias the nanojunctions and to record the current, respectively. The latter was
    measured through a Femto DLPCA200 current amplifier. During subgap measurements the gain of the current amplifier was set to a higher value
    (typically $10^{6}-10^{9}$) and a 1:100 voltage division was applied to increase the signal-to-noise ratio of the voltage bias. The resistive switching
    $I(V)$ curves were measured at $10^{4}$ gain bypassing the voltage divider.}
    \label{fig6}
\end{figure}

The diagram of the electric circuit identically utilized in the low temperature STM and MCBJ measurement setups is shown in Fig.~\ref{fig6}. Three main stages
were utilized for noise filtering, established in a symmetric arrangement: long cryogenic coaxial cables operating as an RC-filter at their full length;
commercially available \emph{MiniCircuits VLFX-80} low pass filters with $f_c=145$~MHz cut-off frequency and 40\,dB insertion loss up to 20\,GHz; and
custom-built RC-filters made of SMD elements ($R=100~\Omega$, $C=2$~nF at $T=1.4$~K). Originating from these filtering elements and the input impedance of the
current amplifier, an $R_S=300~\Omega$ total serial resistance was connected to the sample. The $I(V)$ characteristics are displayed throughout the paper as a
function of the voltage drop on the nanojunction ($V_{\rm bias}$, bias voltage), while triangular signals with  $f_{\rm drive}=2.5$~Hz frequency and $V_{\rm drive}^0$ amplitude are applied (taking the optional 1:100 division into account) by the DAQ unit. The low temperature STM point contact measurements were performed at 1.4~K in a \emph{Janis Research SVT200T-5} liquid helium cryostat.

\subparagraph*{Measurement protocol.}

The distant voltage ranges of resistive switching ($\sim$\,V) and subgap characteristics ($\sim$\,mV) require an automated measurement technique capable of
controlling the voltage division and the gain of the current amplifier simultaneously. While recording $I(V)$ characteristics in the subgap regime, a 1:100
division was applied to the drive voltage, controlled by a relay. In order to prevent degradation of the nanojunctions due to transient voltage spikes, another
relay was used to ground the circuit while switching the gain of the current amplifier.

\subparagraph*{Fitting procedure of subgap $I(V)$ traces.}

The fitting of subgap $I(V)$ characteristics was performed with a type of simulated annealing algorithm using Monte Carlo method written by G. Rubio-Bollinger and co-workers\cite{Rubio-Bollinger-subgap}. This algorithm inputs $I(V)$  traces normalized with $\Delta$ and $G_0$ as demonstrated in Fig.~\ref{fig1}e. The $\Delta$ value is determined form the $I(V)$ traces of the OFF states, which show a tunneling-like characteristic for all the investigated junctions.  In our analysis the fitting is performed in the $0<e\text{V} <6\Delta$ interval using $M=5$ independent conductance channels. This is already enough to resolve the full compexity of the $s$ and $d$ channels in atomic-sized wires\cite{Scheer1998}, but the resolution of more channels would require better voltage resolution. It is emphasized, that in the vicinity of $G_N=1\,$G$_0$ conductance the subgap traces significantly differ if the transport is dominated by a single channel, or if the same conductance is shared between multiple, partially reflecting channels (see black curves in Fig.~\ref{fig4}c). Therefore the fitting very clearly identifies single atom nanowires even in the presence of the discussed $\Gamma$-broadening.  

\section*{Acknowledgements}
This work was supported by the BME-Nanonotechnology FIKP grant of EMMI (BME FIKP-NAT) and the NKFI K119797 and K128534 grants. T.N.T. acknowledges the support
of the UNKP-18-2 and UNKP-19-3 New National Excellence Program of the Ministry of Human Capacities. P.M. acknowledges funding from the Marie Curie and the Bolyai fellowship.
The authors are grateful to G. Rubio-Bollinger for the fitting program based on the theory of multiple Andreev reflections.

\section*{Author Contributions}
T.N.T. performed the measurements and the data analysis. P.M. contributed to the subgap analysis of pure Nb nanojunctions.
M.Cs. contributed to the characterization of the resistive switching properties and the investigation of compositionally symmetric resistive switching
junctions. All authors contributed to the discussion of the results. The manuscript was written by A.H., T.N.T. and M.Cs. The project was supervised by A.H.

\bibliographystyle{achemso}
\bibliography{References}

\clearpage

\begin{figure}[h]
		\centering
		\includegraphics[width=3.25 in]{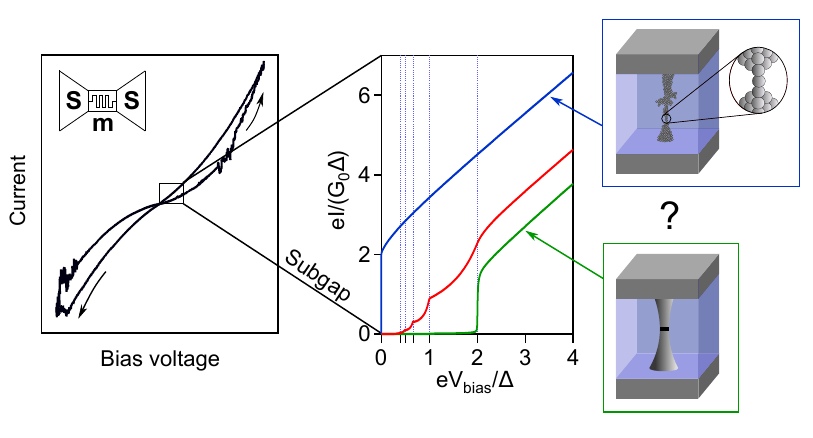}
		\caption{\it \textbf{For Table of Contents Only}}
\end{figure}

\end{document}